\documentstyle[aps]{revtex}

\author{H.Falomir,
R.E.Gamboa Sarav\'\i \  and
E.M.Santangelo
\\ \hfill
\\
Departamento de F\'{\i}sica\\
Facultad de Ciencias Exactas,\\ Universidad Nacional de La Plata, Argentina
}

\title{ Dirac Operator on a disk with global boundary
conditions.\thanks{Partially supported by CONICET, Argentina. } }
\date{October 2, 1996}

\def\dfrac#1#2{{\displaystyle {#1 \over #2}}}
\def\QATOP#1#2{{#1 \atop #2}}
\def\QDATOP#1#2{{\displaystyle {#1 \atop #2}}}

\def\dint{\displaystyle \int }
\def\k{\mbox{\large $\kappa$}}
\def\ka{\mbox{$\kappa$}}
\def\nn{\nonumber}

\begin{document}

\maketitle

\begin{abstract}
We compute the functional determinant for a Dirac operator in the presence of
an Abelian gauge field on a bidimensional disk,  under
global boundary
conditions of the type introduced by Atiyah-Patodi-Singer. We also discuss the
connection between our result and  the index theorem.

\bigskip


\end{abstract}

\bigskip
\bigskip

\section{Introduction}
The wide application of functional determinants in Quantum and Statistical
Physics is by now a well known fact. Typically, one is faced to the necessity
of defining a regularized determinant  for elliptic differential operators,
among which the Dirac first order one plays a central role. An interesting
related problem is the modification of physical quantities due to de presence
of boundaries. The study of boundary effects has lately received much attention
\cite{r0,r1,r2,r3,r4,r5,r6,r7,r8}, since it is of importance in many different
situations,
like effective  models for strong interactions , quantum cosmology  and
application of QFT to statistical systems , among others.

In previous work \cite{a1,a2}, we have studied elliptic Dirac boundary problems
in the case of local boundary conditions. In particular, we have developed for
this case a scheme for evaluating determinants from the knowledge of the
associated Green's function, based on Seeley's theory of complex powers
\cite{Seeley}.

Another type of boundary conditions extensively studied in the literature are
global ones, of the type introduced by Atiyah, Patodi and Singer (APS)
\cite{APS} in connection with the index theorem for manifolds with boundaries
(see  \cite{egh} for a review.)
Other  motivations for considering these global (or spectral) conditions are
their consistency with charge conjugation and chiral invariance, and the
presence of topological obstructions for the chiral Dirac operator under local
boundary conditions (although this restriction no longer holds when considering
the whole Dirac operator \cite{a1}.)

In this paper we present the complete evaluation of the determinant of the
Dirac operator on a disk, in the presence of an axially symmetric Abelian flux,
under spectral boundary conditions (see \cite{Sitenko,Moreno} and references
therein
for related work).

In section II we establish our conventions and set the problem to be
considered.

In Section III we study zero modes and obtain the projector on the null space
of the operator.

Section IV is devoted to the evaluation of the Green's function of the problem
in the subspace orthogonal to such null space. The knowledge of this Green's
function is necessary in the evaluation of the determinant, which is done in
Section V. This  evaluation  is performed in two steps. The
first one involves the consideration of the quotient of determinants with and
without gauge field, under a fixed boundary condition. For this situation, we
grow the  field along a continuous path through a parameter $\alpha$, and
perform a point splitting regularization of the $\alpha$-derivative of the
logarithm of the quotient of determinants, written in terms of the Green's
function.

The second step amounts to the evaluation, via $\zeta$-function, of the
quotient of the free operators, under different global boundary conditions,
which is possible thanks to the explicit determination of their eigenfunctions
and corresponding eigenvalues.

 In Section VI we discuss the application of  the APS index theorem to our
example, with emphasis on the effect of the boundary isomorphism $\sigma$
appearing in the Dirac operator.

Finally, Section VII contains our conclusions.

\section{Setting of the problem}
We will evaluate the
determinant of the operator $D=\ \not \!\!\!i\partial +\not \!\!A$
acting on functions
defined on a two dimensional disk of radius $R,$ under APS boundary conditions.

We take $A_\mu $ to be an Abelian gauge field. As it
is well
known, it can be written as $A_\mu =\epsilon _{\mu \nu }\ \partial
_\nu \phi+\partial_\mu \eta$ $
\ (\epsilon _{01}=-\epsilon _{10}=1)$.  We set $\eta \equiv 0$, thus choosing
the Lorentz gauge. For  $\phi $ we take a smooth
bounded
function $\phi =\phi (r)$; then $A_r=0$ and $A_\theta
(r)=-\partial_r\phi (r)=-\phi^{^{\prime }}(r)$.
We call
\begin{equation}
\label{3.2}{\k } =\dfrac{\Phi}{2\pi}=\dfrac{1}{2\pi}\oint_{r=R}\ A_\theta \ R\
d\theta =- R\phi^{^{\prime }}(R).
\end{equation}

Our convention for two dimensional Dirac matrices is
\begin{equation}
\label{gammas}\gamma _0=\left(
\begin{array}{cc}
0 & 1 \\
1 & 0
\end{array}
\right) ~,\qquad \gamma _1=\left(
\begin{array}{cc}
0 & -i \\
i & 0
\end{array}
\right) ~,\qquad \gamma _5=\left(
\begin{array}{cc}
1 & 0 \\
0 & -1
\end{array}
\right) ,
\end{equation}
which satisfy
\begin{equation}
\gamma _\mu \gamma _\nu \ =\delta _{\mu \nu }\ I\ +i \ \epsilon _{\mu
\nu }\
\gamma _5.
\end{equation}
Therefore, the free Dirac operator can be written as
\begin{equation}
i\not \! \partial =i\ (\gamma _0\ \partial _0+\gamma _1\ \partial
_1)=2i\
\left(
\begin{array}{cc}
0 & \frac \partial {\partial X} \\
\frac \partial {\partial X^{*}} & 0
\end{array}
\right) ,
\end{equation}
where $X=x_0+i\ x_1$ and $X^{*}=x_0-i\ x_1$
or, in polar coordinates
\begin{equation}
\label{Dpolar}
i\not \! \partial =i(\gamma _r\ \partial _r+\frac 1r\gamma _\theta \
\partial
_\theta ),
\end{equation}
with
\begin{equation}
\label{Gpolar}
\gamma _r=e^{-i\gamma _5 \theta}\gamma _0=\left(
\begin{array}{cc}
0 & e^{-i\theta } \\
e^{i\theta } & 0
\end{array}
\right) ,\qquad \gamma _\theta =e^{-i\gamma _5 \theta}\gamma _1=\left(
\begin{array}{cc}
0 & -ie^{-i\theta } \\
ie^{i\theta } & 0
\end{array}
\right) .
\end{equation}

With these conventions, the full Dirac operator can be written as
\begin{equation}
\label{op}D=e^{-\gamma _5\phi (r)\ }i\not \! \partial \ e^{-\gamma
_5\phi (r)}=\left(
\begin{array}{cc}
0 & \sigma ^{-1}(\partial _r+B) \\
\sigma \ (-\partial _r+B) & 0
\end{array}
\right),
\end{equation}
with
\begin{equation}
\sigma =-\ i\ e^{i\theta },
\end{equation}
and
\begin{equation}
B(r)=-\frac ir\ \partial _\theta +\ \partial _r\phi (r).
\end{equation}
At the boundary, the self adjoint operator $B(r)$ becomes
\begin{equation}
B\equiv B(R)=-\frac 1R[\ i\ \partial _\theta+ {\k }],
\end{equation}
with eigenvectors $e^{in\theta }$ and eigenvalues $\frac 1R(n-{\k })$. We take
the radial variable to be conveniently adimensionalized throught multiplication
by a fixed constant with  dimensions of mass.

We will consider the action of the differential operator $D$ on the space of
functions satisfying global boundary conditions, characterized by
\begin{equation}
\label{apsbc}\left(
\begin{array}{cc}
{\cal P}_{\geq} & \sigma (1-{\cal P}_{\geq})\ \sigma ^{*}
\end{array}
\right) \left( \QATOP{\varphi (R,\theta )}{\chi (R,\theta )}\right) =0,
\end{equation}
with
\begin{equation}
{\cal P}_{\geq}=\frac 1{2\pi }\sum_{n\geq k+1}e^{in\theta }\left\langle
e^{in\theta
},\ \cdot \ \right\rangle ,
\end{equation}
where $k$ is the integer such that $k<{\k } \leq k+1$.
Notice that
\begin{equation}
\sigma (1-{\cal P}_{\geq})\sigma^{*}=\sigma \, {\cal P}_{<}\, \sigma^{*}=\frac
1{2\pi }\sum_{n\leq k+1}e^{in\theta }\left\langle e^{in\theta
},\ \cdot \ \right\rangle ={\cal P}_{\leq}
\end{equation}
and the operator so defined, $
\left( D\right) _{{\ka } }$, turns out to be self adjoint. The presence of
$\sigma$  as in (\ref{apsbc}) has a relevant consequence  on the boundary
conditions for the lower components. As we will discuss later, this  also
reflects
on the form of the index theorem for
the present situation.

\bigskip

Our aim is to compute the quotient of the determinants of the operators $
\left( D\right) _{{\ka} }$ and $(i\not \!\! \partial )_{{\ka }=0}.$
Since the global boundary  conditions in Eq. (\ref{apsbc}) depend on the flux
$\Phi $ as a step function,
there is no continuous family connecting both operators. So, we will proceed
in two steps:
\begin{equation}
\left( D\right) _{{\ka } }\rightarrow (i\not \! \partial
)_{{\ka }}\rightarrow (i\not \! \partial  )_{{\ka }=0} .
\end{equation}

In the first step, where  there is no change of boundary conditions, we can
grow the gauge field by varying $\alpha$ from $0$ to $1$ in
\begin{equation}
\label{opalf}D_\alpha =i\not \! \partial +\alpha \not \! \!
A=e^{-\alpha \gamma
_5\phi (r)\;\ }i\not \! \partial \ e^{-\alpha \gamma _5\phi (r)},\
\rm{with }\
0\leq \alpha \leq 1,
\end{equation}
thus going smoothly from the free to the full Dirac
operator. The explicit knowledge of the Green's function will allow us to
perform the calculation of this step, where we will use a gauge invariant point
splitting regularization of the $\alpha$-derivative of the determinant. The
second step will be achieved by using a $\zeta$-function regularization, after
explicitly computing the spectra.

But this is not the whole story: As we are
going to see, these global boundary conditions give rise to  the presence of
zero modes, which   must be taken  into account.

\section{Zero modes}

We will here show that the operator $(D_\alpha )_{{\ka }}$ has $\vert
k+1\vert $ zero modes.

{}From (\ref{op}), we get
\begin{equation}
D_\alpha \ \psi =0\Rightarrow \not \! \partial \ e^{-\alpha \gamma
_5\phi (r)}\
\psi =0,
\end{equation}
or, equivalently
\begin{equation}
\left(
\begin{array}{cc}
0 & e^{-i\theta }(\partial _r-\frac ir\partial _\theta ) \\
e^{i\theta }(\partial _r+\frac ir\partial _\theta ) & 0
\end{array}
\right) \left(
\begin{array}{cc}
e^{-\alpha \phi (r)} & 0 \\
0 & e^{\alpha \phi (r)}
\end{array}
\right) \left( \QATOP{\varphi (r,\theta )}{\chi (r,\theta )}\right)
=0.
\end{equation}
Now, we introduce the expansions
\begin{eqnarray}
\displaystyle \varphi (r,\theta )=\sum\limits_{n=-\infty }^\infty
\varphi _n(r)\
e^{in\theta }, \nn  \\
\displaystyle \chi (r,\theta )=\sum\limits_{n=-\infty }^\infty \chi
_n(r)\ e^{in\theta }.
\nn
\end{eqnarray}
The solutions are thus given by
\begin{eqnarray}
\displaystyle \varphi _n(r)=\ a_{n\ }\ r^n\ e^{\alpha \phi (r)},
\nn \\ \displaystyle \chi _n(r)=\ b_{n\ }\ r^{-n}\ e^{-\alpha \phi (r)},
\end{eqnarray}
where the coefficients $a_n$ and $b_n$ are to be determined from the
normalizability requirement at the origin and the
boundary conditions at  $r=R$, Eq. (\ref{apsbc}). Thus
\begin{eqnarray}
\nn
\varphi (r,\theta )=\ e^{\alpha \phi (r)}\sum\limits_{n=0}^k\, a_{n\ }r^n\
e^{in\theta }, \\
\chi (r,\theta )=\ e^{-\alpha \phi (r)}\sum\limits_{n=0}^{-k-2} \,
b^{\prime}_{n\ }\
r^{n}\ e^{-in\theta }.
\end{eqnarray}
Then, there are $\vert k+1\vert $ zero modes which, once normalized, are
given by
\begin{eqnarray}
\nn
\dfrac{\ e^{\alpha \phi (r)}}{\sqrt{2\pi \ q_n(R;\alpha)}}\ \left(
\QDATOP{X^n}{0}
\right) ,{\rm \ \   for }\ 0\leq n\leq k,{\rm \  if }\ \ k\geq 0; \\
\nn  \\ \nn\\
\dfrac{\ e^{-\alpha \phi (r)}}{\sqrt{2\pi \ p_n(R;\alpha)}}\ \left( \QDATOP{0}{
X^{*n}}\right) \ ,{\rm\ \ for }\ \ 0\leq n\leq -k-2,{\rm\  if }\ \ k<-1.
\end{eqnarray}
Here, the normalization factors are
\begin{eqnarray}
\label{qene}
\nn
q_n(u;\alpha)=\dint_0^u\ e^{2\alpha \phi (r)}\ r^{2n+1}\ dr\ , \\
p_n(u;\alpha)=\dint_0^u\ e^{-2\alpha \phi (r)}\ r^{2n+1}\ dr\ .
\end{eqnarray}
Notice that, for $k=-1$ (in particular, when $\Phi =0)$, there is no zero mode.

So, the kernel of the orthogonal projector on Ker$(D_\alpha )$ is

\begin{eqnarray}
P_\alpha (z,w)=\sum\limits_{n=0}^k\dfrac{\ e^{\alpha [\phi (z)+\phi (w)]}}{
2\pi \ q_n(R;\alpha)}\left(
\begin{array}{cc}
(ZW^{*})^n & 0 \\
0 & 0
\end{array}
\right) ,
{\rm if }\ k\geq 0,  \nn \\  \nn\\
P_\alpha (z,w)=\sum\limits_{n=0}^{-k-2}\dfrac{\ e^{-\alpha [\phi (z)+\phi
(w)]}}{2\pi \ p_n(R;\alpha)}\left(
\begin{array}{cc}
0 & 0 \\
0 & (Z^{*}W)^n
\end{array}
\right) ,{\rm\  if }\ k<-1.
\end{eqnarray}

Since $ P_\alpha$ is an orthogonal projector, $P_\alpha^2=P_\alpha $, we have
\begin{equation}
\label{proj}
\partial _{\alpha}P_\alpha\, \, \, (1-P_\alpha)=P_\alpha \, \, \, \partial
_{\alpha}P_\alpha.
\end{equation}

\bigskip

Being  $(D_\alpha +P_\alpha )_{{\ka }}$  invertible, we can define
\begin{equation}
Det^{^{\prime }}(D_\alpha )_{{\ka }}\equiv Det(D_\alpha +P_\alpha
)_{{\ka }},
\end{equation}
and write
\begin{equation}
\label{dobleco}\frac{Det^{^{\prime }}(D)_{{\ka }}}{Det(i\not \! \partial
)_{{\ka }=0}}=\frac{Det(D+P_1)_{{\ka }}}{Det(i\not \! \partial
+P_0)_{{\ka }}}\ \frac{Det(i\not \! \partial  +P_0)_{{\ka }}}{Det(i\not \!
\partial  )_{{\ka }=0}}.
\end{equation}

\bigskip

As mentioned above, we can compute
\begin{equation}
\label{dalfa}
\dfrac \partial {\partial \alpha }\left[\ln Det{(D_{\alpha} +P_{\alpha})_{{\ka
}}}\right]=Tr \left[ (\not \! \! A+ \partial_{\alpha}P_{\alpha} ) G_\alpha
\right],
\end{equation}
after determining the  Green's function
$G(x,y)$ satisfying
\begin{eqnarray}
(D_\alpha +P_\alpha )\ G(x,y)=\delta (x,y), \nn
\\
\left(
\begin{array}{cc}
{\cal P}_{\geq} & \, \,  {\cal P}_{\leq}
\end{array}
\right) G(x,y)\vert _{r=R }=0,
\end{eqnarray}
and applying a regularization prescription to the trace in (\ref{dalfa}). Then,
by integrating in $\alpha$ from $0$ to $1$ we will get the first  quotient in
the r.h.s. of (\ref{dobleco}).

The computation of this Green's function will be the subject of the next
section.

\section{Calculation of the Green's function}

Notice that, since $D_\alpha$ is self adjoint,  we can write
\begin{equation}
\label{green2}G(x,y)=(1-P_\alpha )\ {\cal G}(x,y)\ (1-P_\alpha )+\ P_\alpha,
\end{equation}
where ${\cal G}(x,y)$ is the kernel of the right-inverse of $D_\alpha $ on
the orthogonal complement of $Ker(D_\alpha ).$ From the Eq. (\ref{op}), it is
easy to see that
\begin{equation}
{\cal G}(x,y)=e^{\alpha \gamma _5\phi (r)}\ {\cal G}_0(x,y)\ e^{\alpha
\gamma _5\phi (r^{\prime} )},
\end{equation}
where ${\cal G}_0(x,y)$ can be obtained by solving the differential equation
\begin{equation}
i\not \! \partial  \left( \QATOP{\varphi }{\chi }\right) =\left( \QATOP{f}{g}
\right) ,
\end{equation}
for $\left( \QATOP{\varphi }{\chi }\right) $ satisfying
\begin{eqnarray}
P_\alpha \ \ e^{\alpha \gamma _5\phi (r )}\ \left(
\QDATOP{\varphi }{\chi }\right) =0, \nn  \\
\nn \\
\left(
\begin{array}{cc}
{\cal P}_{\geq} & \, \,  {\cal P}_{\leq}
\end{array}
\right)  e^{\alpha \gamma _5\phi (R )} \left( \QDATOP{\varphi }{\chi }\right)
=0.
\end{eqnarray}
Now, by expanding
\begin{eqnarray}
\varphi (r,\theta )=\sum \varphi _n(r)\ e^{in\theta },\, \, \, \chi (r,\theta
)=\sum \chi _n(r)\ e^{in\theta } \nn  \\  \nn\\
f(r,\theta )=\sum f_n(r)\ e^{in\theta },\ g(r,\theta )=\sum g_n(r)\
e^{in\theta },
\end{eqnarray}
we get
\begin{eqnarray}
\varphi (r,\theta )=i\sum\limits_{n\geq k+1}e^{in\theta
}r^n\int_r^Rdr^{\prime }\ (r^{\prime })^{-n}\ g_{n+1}(r^{\prime })
-\ i\sum\limits_{n\leq k}e^{in\theta }r^n\int_0^rdr^{\prime }\
(r^{\prime })^{-n}\ g_{n+1}(r^{\prime }) \nn\\
\nn\\
=\frac i{2\pi }\int_r^Rdr^{\prime }\int_0^{2\pi }d\theta ^{\prime
}\sum\limits_{n\geq k+1}e^{in(\theta -\theta ^{\prime })}\left( \frac
r{r^{\prime }}\right) ^ne^{-i\theta ^{\prime }}g(r^{\prime },\theta ^{\prime
})
-\frac i{2\pi }\int_0^rdr^{\prime }\int_0^{2\pi }d\theta ^{\prime
}\sum\limits_{n\leq k}e^{in(\theta -\theta ^{\prime })}\left( \frac
r{r^{\prime }}\right) ^ne^{-i\theta ^{\prime }}g(r^{\prime },\theta ^{\prime
}).
\end{eqnarray}
Analogously
\begin{equation}
\chi (r,\theta )=\frac i{2\pi }\int_r^Rdr^{\prime }\int_0^{2\pi }d\theta
^{\prime }\sum\limits_{n\leq k+1}e^{in(\theta -\theta ^{\prime })}\left(
\frac{r^{\prime }}r\right) ^ne^{i\theta ^{\prime }}f(r^{\prime },\theta
^{\prime })
-\frac i{2\pi }\int_0^rdr^{\prime }\int_0^{2\pi }d\theta ^{\prime
}\sum\limits_{n\geq k+2}e^{in(\theta -\theta ^{\prime })}\left( \frac{%
r^{\prime }}r\right) ^ne^{i\theta ^{\prime }}f(r^{\prime },\theta ^{\prime
}).
\end{equation}
Therefore, we  have
\begin{equation}
{\cal G}_0(x,y)=\frac 1{2\pi i}\left(
\begin{array}{cc}
0 & \frac 1{X-Y}\left( \frac XY\right) ^{k+1} \\
\frac 1{X^{*}-Y^{*}}\left( \frac{Y^{*}}{X^{*}}\right) ^{k+1} & 0
\end{array}
\right) ,
\end{equation}
and
\begin{equation}
{\cal G}(x,y)=\frac 1{2\pi i}\left(
\begin{array}{cc}
0 & \frac{e^{\alpha [\phi (x)-\phi (y)]}}{X-Y}\left( \frac XY\right) ^{k+1}
\\ \frac{e^{-\alpha [\phi (x)-\phi (y)]}}{X^{*}-Y^{*}}\left( \frac{Y^{*}}{%
X^{*}}\right) ^{k+1} & 0
\end{array}
\right) .
\end{equation}
Finally, from Eq. (\ref{green2}), after a straightforward but tedious
computation, we obtain, for  $ k \geq 0$,
\begin{equation}
\label{fgaps}
G(x,y)=
\dfrac 1{2\pi i}\left( \
\QDATOP{ie^{\alpha [\phi (x)+\phi (y)]}\sum\limits_{n=0}^k\frac{(XY^{*})^n}{%
q_{n(R;\alpha)}}\qquad \qquad \ \ \qquad e^{\alpha [\phi (x)-\phi (y)]}\left[
\frac
1{X-Y}+\frac 1Y\sum\limits_{n=0}^k\left( \frac XY\right)
^n\frac{q_{n(y;\alpha)}}{%
q_{n(R;\alpha)}}\right] }{e^{-\alpha [\phi (x)-\phi (y)]}\left[ \frac
1{X^{*}-Y^{*}}-\frac 1{X^{*}}\sum\limits_{n=0}^k\left( \frac{Y^{*}}{X^{*}}%
\right) ^n\frac{q_{n(x;\alpha)}}{q_{n(R;\alpha)}}\right] \ \ \qquad \qquad
\qquad \qquad
0\ \ \ \ \ \ \ \ \qquad \ }\right) ,
\end{equation}
and,  for $k<-1$,
\begin{equation}
\label{fgkn}
G(x,y)=
\dfrac 1{2\pi i}\left(
\QDATOP{\qquad \qquad \qquad \qquad 0\qquad \qquad \qquad \qquad e^{\alpha
[\phi (x)-\phi (y)]}\left[ \frac 1{X-Y}-\frac
1X\sum\limits_{n=0}^{-k-2}\left( \frac YX\right)
^n\frac{p_{n(x;\alpha)}}{p_{n(R;\alpha)}}%
\right] }{e^{-\alpha [\phi (x)-\phi (y)]}\left[ \frac 1{X^{*}-Y^{*}}+\frac
1{Y^{*}}\sum\limits_{n=0}^{-k-2}\left( \frac{X^{*}}{Y^{*}}\right) ^n\frac{%
p_{n(y;\alpha)}}{p_{n(R;\alpha)}}\right] \ \ \qquad \ \ \ ie^{-\alpha [\phi
(x)+\phi
(y)]}\sum\limits_{n=0}^{-k-2}\frac{(X^{*}Y)^n}{p_{n(R;\alpha)}}}\right) .
\end{equation}

\section{Evaluation of the determinant}

Now  we have all the necessary elements to compute the first quotient in the
r.h.s. of Eq. (\ref{dobleco}). In fact,  from Eqs. (\ref{green2}) and
(\ref{proj})

\begin{equation}\
Tr \left[ (\partial_\alpha P_\alpha )\, G_\alpha \right]=0.
\end{equation}
So, Eq. (\ref{dalfa}) becomes

\begin{equation}
\label{dalfa2}
\dfrac \partial {\partial \alpha }\left[\ln Det{(D_{\alpha} +P_{\alpha})}_{
{\ka }} \right]=Tr \left[ \not  \! \! A G_\alpha \right].
\end{equation}
As usual, the kernel of the operator inside the trace is singular at the
diagonal, so we must introduce a regularization. We will employ a
point-splitting one where, following Schwinger \cite{Schwinger}, we will
introduce a  phase factor  in order to preserve gauge invariance. We thus get,
for $k \geq 0$, from (\ref{fgaps})
\begin{eqnarray}
Tr \left[ \not  \! \! A G_\alpha \right]=\int_{r<R} d^2x \, \,  tr\left[ \not
\! \! A(x) G_\alpha(x,x+\epsilon) e^{i \alpha \epsilon \cdot A(x)}
\right]_{\epsilon \rightarrow 0} \nn\\    \nn  \\  \nn
=\displaystyle -\frac 1{2\pi }\int_{r<R} d^2x  \, \phi^{\prime}(r)
\left\{  e^{ -i \theta}  e^{ \alpha \phi^{\prime}(r) [\epsilon_r -i
\epsilon_\theta]} \left[   \frac{e^{i\theta}}{\epsilon_r -i \epsilon_\theta}+
\frac{1}{X^{*}}\sum\limits_{n=0}^k
\frac{q_{n(x;\alpha)}}{q_{n(R;\alpha)}}+O(\epsilon)
\right]  \right.
\\ \nn \\ \nn
\displaystyle +\left. e^{ i \theta}  e^{ -\alpha \phi^{\prime}(r) [\epsilon_r
+i \epsilon_\theta]} \left[   -\frac{e^{-i\theta}}{\epsilon_r +i
\epsilon_\theta}+
\frac{1}{X}\sum\limits_{n=0}^k
\frac{q_{n(x;\alpha)}}{q_{n(R;\alpha)}}+O(\epsilon)
\right]
 \right\}_{\epsilon \rightarrow 0}
 \\ \nn \\ \nn
=\displaystyle -\left[\frac{2i \epsilon_\theta}{\epsilon_r^2
+\epsilon_\theta^2}\right]_{\epsilon \rightarrow 0} \, \int_0^R dr \, r
\phi^{\prime}(r)
-\frac \alpha \pi \int_{r<R} d^2x\
{\phi ^{\prime }}^2
\displaystyle -2\ (k+1)\ \phi (R)\ +\sum\limits_{n=0}^k\ \dfrac \partial
{\partial \alpha }\ln \left[q_n(R;\alpha)\right].\\
\end{eqnarray}

Finally, performing a ``symmetric" $\epsilon \rightarrow 0$ limit, which drops
out the first term, and  integrating in $\alpha $ from $0$ to $1$, we get
\begin{equation}\label{a1}
\ln
\left[
\frac{Det(D+P_1)_{{\ka }}}{Det(i\not \! \partial +P_0)_{{\ka }}}
\right] =-\frac
1{2\pi }\int _{r<R} d^2x\  {\phi ^{\prime }}^2 -2\ (k+1)\ \phi (R)\
+\sum\limits_{n=0}^k\
\ln \left[2(n+1)\frac{q_n(R;1)}{R^{2(n+1)}}\right]\ .
\end{equation}

When there are no zero modes $(k+1=0)$ only the first term in the r.h.s.
survives. For $k<-1$, from (\ref{fgkn}), one gets a similar expression where
the sum   runs from $0$ to $-k-2= \vert k+1 \vert -1$, and the norms, defined
in Eq. (\ref{qene}), $q_n$, are replaced by $p_n$.  This result can also be
obtained by performing the change $\phi \rightarrow -\phi$ and $k+1 \rightarrow
-(k+1)$. Moreover, as expected from (\ref{op}),  this expression is invariant
under the shift $\phi(r)  \rightarrow \phi(r)+\epsilon $, for any constant
$\epsilon$.

\bigskip

In the following, we will obtain the second quotient of determinants in
Eq. (\ref{dobleco}) by computing explicitly the spectra of the free Dirac
operators and using a $\zeta$-function regularization. The
equation to be solved is
\begin{equation}
i\not \! \partial  \psi =\lambda \psi ,
\end{equation}
or, explicitly
\begin{equation}
\label{eqdif}
i \left(
\begin{array}{cc}
0 & e^{-i\theta }(\partial _r-\frac ir\partial _\theta ) \\
e^{i\theta }(\partial _r+\frac ir\partial _\theta ) & 0
\end{array}
\right) \left( \QATOP{\varphi (r,\theta )}{\chi (r,\theta )}\right) =\lambda
\ \left( \QATOP{\varphi (r,\theta )}{\chi (r,\theta )}\right) .
\end{equation}
Now, by expanding
\begin{eqnarray}
\varphi (r,\theta )=\sum\limits_{n=-\infty }^\infty \varphi _n(r)\
e^{in\theta }, \nn \\
\\ \nn
\chi (r,\theta )=\sum\limits_{n=-\infty }^\infty \chi _n(r)\ e^{in\theta },
\end{eqnarray}
we get
\begin{eqnarray}
i\ \varphi _n^{\prime }(z)-i\ \dfrac nz\varphi _n(z)=\frac \lambda {\vert
\lambda \vert }\ \chi _{n+1}(z) \nn \\
\\ \nn
i\ \chi _{n+1}^{\prime }(z)+i\ \dfrac{(n+1)}z\ \chi _{n+1}(z)=\frac \lambda
{\vert
\lambda \vert }\ \varphi _n(z)
\end{eqnarray}
where $z=\vert \lambda \vert r.$ So, $\varphi _n(z)$ satisfies
\begin{equation}
\ \varphi _n^{\prime \prime }(z)+\ \dfrac 1z\varphi _n^{\prime }(z)+(1-\frac{
n^2}{z^2})\varphi _n(z)=0.
\end{equation}

Therefore, the general solution of Eq. (\ref{eqdif}) is
\begin{equation}
\psi (r,H)=\sum_{n=0}^\infty \psi _n\left( \QATOP{J_n(\vert \lambda \vert r)\
e^{in\theta }}{-i\frac{\vert \lambda \vert} { \lambda }J_{n+1}(\vert \lambda
\vert
r)\ e^{i(n+1)\theta }}\right) .
\end{equation}

Now, by imposing the global boundary conditions  given in Eq. (\ref{apsbc}), it
follows that: If $n\geq k+1$ , the upper component must vanish at the
boundary, which implies $\lambda =\pm j_{n,l}/R$  ( $j_{n,l}$ is
the $l$-th zero of $J_n(z)$). Analogously, if $n\leq k,$ it is the lower
component which must vanish, and so $\lambda =\pm j_{n+1,l}/R.$ Therefore,
the eigenvalues are
\begin{equation}
\lambda _{n,l}=\pm j_{n,l}/R,\ {\rm for\ }n=0,\pm 1,\pm 2,...\ {\rm and\ }\ \
l=1,2,...\ .
\end{equation}
Notice that $j_{-n,l}=j_{n,l} $, and that, for $n=k+1$ the eigenvalues appear
twice, once for an eigenfunction with vanishing upper component at the
boundary,  and once for  another one with vanishing lower component.

{}From these eigenvalues we can construct the $\zeta $-function
\begin{equation}
\label{zeta}
\displaystyle \zeta _{(i\not\partial  +P_0)_{{\ka }}}(s)=\vert k+1\vert
+(1+e^{-i\pi s})\
\left\{ \sum\limits_{n=-\infty }^{n=\infty }\sum\limits_{l=1}^\infty \ \left(
\dfrac{
j_{n,l}}R\right) ^{-s}+\sum\limits_{l=1}^\infty \ \left( \dfrac{j_{\vert
k+1\vert,l}}%
R\right) ^{-s}\right\} .
\end{equation}
The first term, $\vert k+1\vert $, is the multiplicity of the 0-eigenvalue of
$(i\not \! \partial )_{{\ka }}$. It is also interesting to note that the double
sum in the r.h.s. corresponds to the $\zeta$-function of the Laplacian on a
disk with Dirichlet
boundary conditions, thus being  analytic at $s=0$ \cite{Seeley} .
The second sum, to be explicitly computed below, is regular at $s=0$. Then
$\zeta _{(i\not\partial  +P_0)_{{\ka }}}(s)$ is regular at the origin. As far
us we know, the $\zeta$-regularity at the origin for non local boundary
conditions has not been established in general \cite{SG}.

{}From (\ref{zeta}), we can write
\begin{eqnarray}
\label{qdl}
\ln \left[
\dfrac{Det(i\not \! \partial  +P_0)_{\ka }}{Det(i\not \! \partial  )_{{\ka }
=0}}
\right] =-\dfrac d{ds}\left[ \zeta _{(i\not\partial  +P_0)_{{\ka }}}(s)-\zeta
_{(i\not\partial  )_{{\ka=0 }}}(s)\right] _{s=0} \nn \\  \nn \\
=-\dfrac d{ds}\left[ (1+e^{-i\pi s})\ R^s\left\{
\sum\limits_{l=1}^\infty (\ j_{\vert
k+1\vert,l})^{-s}-\sum\limits_{l=1}^\infty \
(j_{0,l})^{-s}\right\} \right] _{s=0}  \nn \\ \nn \\
=-2\left[ f_{\vert k+1\vert}^{\prime }(0)-f_0^{\prime }(0)+(\ln R-\frac{i\pi }
2)[f_{\vert k+1\vert}(0)-f_0(0)]\right] ,
\end{eqnarray}
where the function $f$ is defined as
\begin{equation}
f_\nu (s)\equiv \sum_{l=1}^\infty \ (\ j_{\nu,l})^{-s}.
\end{equation}
Taking into account the asymptotic expansion for the zeros of Bessel
functions \cite{Tabla}
\begin{equation}
(\ j_{\nu ,l})^{-s}-(l\pi )^{-s}+s\left( \frac{2\nu -1}4\right) \pi (l\pi
)^{-s-1}\sim O(l^{-s-2}),
\end{equation}
we can write
\begin{equation}
f_\nu (s)=\sum\limits_{l=1}^\infty \ \left[ (\ j_{\nu ,l})^{-s}-(l\pi
)^{-s}+s\left(
\frac{2\nu -1}4\right) \pi (l\pi )^{-s-1}\right]
+\ \pi ^{-s}\left[ \zeta (s)-s\left( \frac{2\nu -1}4\right) \zeta
(s+1)\right] ,
\end{equation}
where $\zeta (s)$ is Riemann's $\zeta $-function. Notice that the first
series converges for $Re\ s>-1$; thus, it can be evaluated at $s=0$,
to obtain
\begin{equation}
\label{f0}f_\nu (0)=-\frac \nu 2-\frac 14,
\end{equation}
and
\begin{equation}
\label{f'0}f_\nu ^{\prime }(0)=-\frac 12\ln 2+\left( \frac{2\nu -1}4\right)
(\ln \pi -\gamma )-\sum\limits_{l=1}^\infty \ln \left[ \frac{\ j_{\nu ,l}}{%
l\pi }\ e^{-\left( \frac{2\nu -1}{4\ l}\right) }\right] ,
\end{equation}
where $\gamma $ is Euler's constant. By replacing Eqs. (\ref{f0}) and (\ref
{f'0}) into Eq. (\ref{qdl}), we get
\begin{equation}
\label{a2}
\ln \left[ \dfrac{Det(i\not \! \partial  +P_0)_{{\ka }}}{Det(i\not \! \partial
)_{{\ka }=0}}\right] =-\vert k+1\vert [\frac{i\pi }2-\gamma -\ln (\frac R\pi
)]+2\sum\limits_{l=1}^\infty \ln \left[ \frac{\ j_{\vert k+1\vert,l}}{\
j_{0,l}}\
e^{-\left( \frac{\vert k+1\vert}{2\ l}\right) }\right] .
\end{equation}

\bigskip

Finally, adding (\ref{a1}) and (\ref{a2}), and taking into account that we have
used a gauge invariant procedure, we obtain for the quotient in the l.h.s. of
(\ref{dobleco})
\begin{eqnarray}
\label{final}
\displaystyle \ln\left[
\frac{Det(D+P_1)_{{\ka }}}{Det(i\not \! \partial )_{{\ka }=0}}
\right] =-\frac
1{2\pi }\int_{r<R} d^2x\ A_\mu (\delta_{\mu \nu}-
\dfrac{\partial_{\mu}\partial_{\nu}}{\partial^2}) \ A_\nu -2\ (k+1)\ \phi
(R)\nn
\\  \nn \\ \displaystyle +\sum\limits_{n=0}^k\
\ln\left[2(n+1)\frac{q_n(R;1)}{R^{2(n+1)}} \right]
 -\vert k+1\vert [\frac{i\pi }2-\gamma -\ln (\frac R\pi
)]+2\sum\limits_{l=1}^\infty \ln \left[ \frac{\ j_{\vert k+1\vert,l}}{\
j_{0,l}}\
e^{-\left( \frac{\vert k+1\vert}{2\ l}\right) }\right] .
\end{eqnarray}
The first term is the integral on the disk of the same expression appearing in
the well-known result for the boundaryless case \cite{Annals}.
Note that this final result is also invariant under the transformation: $\phi
\rightarrow \phi +\epsilon$ for any constant $\epsilon$.

\bigskip

Now, a comment is in order concerning global axial transformations and their
relationship to zero modes. Under such a transformation, with constant
$\epsilon$,
\begin{equation}
e^{-\gamma_{5} \epsilon }  (i\not \! \partial )_{{\ka }=0}e^{-\gamma_{5}
\epsilon }=(i\not \! \partial )_{{\ka }=0}
\end{equation}
is invariant. Moreover
\begin{equation}
e^{-\gamma_{5} \epsilon }(D+P_1)_{{\ka }}e^{-\gamma_{5} \epsilon }
=(D+e^{-\gamma_{5} \epsilon }P_1e^{-\gamma_{5} \epsilon })_{{\ka }},
\end{equation}
while  its inverse has the structure
\begin{equation}\label{green3}
G^{(\epsilon)}(x,y)=(1-P_1 )\ {\cal G}(x,y)\ (1-P_1 )+e^{\gamma_{5} \epsilon }\
P_1\,  e^{\gamma_{5} \epsilon }.
\end{equation}

Therefore, since $\gamma_5$ leaves Ker $D$ invariant,
\begin{equation}
\label{dep}
\displaystyle \dfrac{\partial}{\partial \epsilon}\ln\left[
\frac{Det\left(e^{-\gamma_{5} \epsilon }(D+P_1)_{{\ka }}e^{-\gamma_{5} \epsilon
}\right)}
{Det\left( e^{-\gamma_{5} \epsilon }(i\not \! \partial )_{{\ka
}=0}e^{-\gamma_{5} \epsilon }\right) }
\right] = -Tr\left[ e^{-\gamma_{5} \epsilon } \left\{ \gamma_5 , P_1\right\}
e^{-\gamma_{5} \epsilon } G^{(\epsilon)} \right]
= -2Tr \left[ \gamma _{5} P_1  \right]=-2(N_{+}-N_{-}),
\end{equation}
where  $N_{+(-)}$ is the number of positive(negative) chirality zero modes.

It can be verified that our strategy leads to the correct result for the index
of $D$. By following the same procedure that  lead to Eq. (\ref{final}), we can
compute the quotient of determinants in the l.h.s of (\ref{dep}). In fact,
using Eq. (\ref{green3}
) instead of Eq. (\ref{green2}), the only difference appears in the first term
of the r.h.s. of (\ref{zeta}), where a factor $e^{\pm 2\epsilon s}$ arises.
Thus, after performing the $\epsilon$-derivative
\begin{equation}
\label{enemas}
N_{+}  -N_{-}= k+1,
\end{equation}
which agrees with our previous result for the number of zero modes.

For the sake of completeness, in the next section we will show how the presence
of the $\sigma$'s in (\ref{op}) and (\ref{apsbc}) affects the index theorem.

\section{Relation to the Index Theorem}

Here we will reobtain the index by following the  Atiyah-Patodi-Singer approach
\cite{APS}.

The Dirac operator in polar coordinates can be written as \cite{Sitenko,Moreno}
\begin{equation}
\label{op1}
D=\dfrac{1}{\sqrt{r}}\left(
\begin{array}{cc}
0 &   e^{-i\theta/2\ }(\partial _r+B)  e^{-i\theta/2}\\
 e^{i\theta/2\ }(-\partial _r+B)  e^{i\theta/2}& 0
\end{array}
\right)\sqrt{r},
\end{equation}
subject to the boundary conditions in Eq. (\ref{apsbc}). It is to be stressed
that the presence of the factors $e^{\pm i\theta/2} $ coming from the $\sigma
$'s will produce a departure from the index formula in \cite{APS}.

To be be consistent with (\ref{apsbc}) we call $\cal D$ the differential
operator $(-\partial _r+B)$ acting  on functions
\begin{equation}
\displaystyle f(r,\theta)= \sum_{i}^{n}{f_{n}(r) e^{i(n+\frac{1}{2})\theta }}
\end{equation}   satisfying
\begin{equation}
{\cal P}_{\geq}   \left[ e^{-i \theta /2}
f(R,\theta) \right]
 =0.
\end{equation}
Its adjoint, ${\cal D}^{+}$, is the differential operator $(\partial _r+B)$
defined on the space of functions
\begin{equation}
g(r,\theta)= \sum_{i}^{n}{g_{n}(r) e^{i(n-\frac{1}{2})\theta }}
\end{equation}
satisfying
\begin{equation}
{\cal P}_{\leq}   \left[ e^{i \theta /2}
g(R,\theta) \right]
 =0.
\end{equation}
Then $ D^+ D= \dfrac{1}{\sqrt{r}} e^{-i \theta /2}\, {\cal D}^+ {\cal D} \,e^{i
\theta /2} {\sqrt{r}}$, where $ {\cal D}^+ {\cal D}= -\partial_{r}^{2}+B^2$
acting on functions satisfying
\begin{equation}
\left\{
\begin{array}{c}
{\cal P}_{\geq}   \left[ e^{-i \theta /2}
f(R,\theta) \right]
 =0 \\ \\
{\cal P}_{\leq}   \left[ e^{i \theta /2}
{\cal D}f(R,\theta) \right]
 =0.
\end{array}
\right.
\end{equation}
or, equivalently,
\begin{equation}
\left\{
\begin{array}{cr}
f_{n}(R)=0 &{\rm for}\, \, n \geq k+1\\ \\
-f_{n}^{\prime }(R) + \dfrac{n+1/2-{\k }}{R} f_{n}(R)=0& {\rm for}\, \, n \leq
k.
\end{array}
\right.
\end{equation}

Similarly, $ {\cal D} {\cal D}^+= -\partial_{r}^{2}+B^2$ acting on functions
$g(r,\theta)$ satisfying
\begin{equation}
\left\{
\begin{array}{cr}
g_{n}(R)=0 &{\rm for}\, \, n \leq k+1\\ \\
g_{n}^{\prime }(R) + \dfrac{n+1/2-{\k }}{R} g_{n}(R)=0& {\rm for}\, \, n \geq
k+2.
\end{array}
\right.
\end{equation}

Now, we have for the heat kernels
\begin{equation}
\label{heat}
Tr\left\{ e^{-t D^+ D}\right\}-Tr\left\{ e^{-t D D^+}\right\}=Tr\left\{ e^{-t
{\cal D^+ D}}\right\}-Tr\left\{ e^{-t {\cal D D}^+}\right\}.
\end{equation}

When written in this fashion, the boundary contribution to the r.h.s.  in
(\ref{heat}), $K(t)$,  can be easily constructed from Eqs. (2.16) and (2.17) in
\cite{APS}, thus getting
\begin{equation}\label{K}
K(t)
= \displaystyle \dfrac {1}{2} \,   {\rm erfc}\left( - ( k+\frac{1}{2}-{\k } )
\sqrt{t}\,   \right)
-\sum_{n \neq k}{ \dfrac{sig(n+\frac{1}{2}-{\k })}{2} \,  {\rm erfc} \left(
\vert  \, n+\frac{1}{2}-{\k } \vert \,  \sqrt{t}\,  \right) }.
\end{equation}
Taking into account that
\begin{equation}\label{m}
\lim_{t\rightarrow \infty}{K(t)}=m({\k })=
\cases{
0 & for\qquad \hfill $k+\frac{1}{2}<{\k } \leq k+1$ \hfill \cr
\frac{1}{2} & for \qquad\hfill ${\k }=k+\frac{1}{2}$ \hfill ,\cr
1 & for \qquad \hfill $k<{\k }<k+\frac{1}{2}$ \hfill
}
\end{equation}
we obtain
\begin{equation}
\int_{0}^{\infty} dt \, \,  t^{s-1} \left\{K(t)-m({\k })\right\} = \frac{-
\Gamma (s+1/2)}{2 s \sqrt{\pi}} \, \, \eta_{(B+\frac{1}{2 R}) }(2 s),
\end{equation}
where
\begin{equation}
\eta_{(B+\frac{1}{2R})}(s)= R^s \sum_{n \neq {\k }-\frac{1}{2}} {
sig(n+\frac{1}{2}-{\k } ) \, \,  \vert n+\frac{1}{2}-{\k } \vert^{-s}}.
\end{equation}
In particular
\begin{equation}\label{eta}
\eta_{(B+\frac{1}{2R})}(0)=
\cases{
2({\k } - k-1) & for\qquad \hfill $k+\frac{1}{2}<{\k } \leq k+1$ \hfill \cr
\hfill 0 \hfill & for \qquad\hfill ${\k }=k+\frac{1}{2}$ \hfill \cr
\hfill2({\k } - k)\hfill & for \qquad \hfill $k<{\k }<k+\frac{1}{2}$ \hfill
}.
\end{equation}
Finally, taking into account that, for $k<{\k }\leq k+1$,
\begin{equation}
m({\k }) -\dfrac{1}{2}\, \eta_{(B+\frac{1}{2R})}(0)=k+1-{\k }
= \dfrac{1}{2} \left[ 1- { h}(B)- \eta_{(B)}(0)\right],
\end{equation}
 where $h(B)=$ dim ker$(B)$, and from the asymptotic expansion \cite{APS} of
the heat
kernels in (\ref{heat}), we have
\begin{equation}
{\rm index}\,  D= {\k } + \frac{ \left[ 1- { h}(B)-
\eta_{(B)}(0)\right]}{2}=k+1,
\end{equation}
in agreement with (\ref{enemas}). The first term in the intermediate expression
is the well known contribution from the bulk \cite{Annals}. The second one is
the boundary contribution of APS, shifted by 1/2. This correction, due to the
presence of the factor $\sigma$ in (\ref{op}), has already been obtained in
\cite{Sitenko}  with slightly different spectral boundary conditions.

\section{Conclusions}

In this paper we have achieved the complete evaluation of the determinant of
the Dirac operator on a disk, in the presence of an axially symmetric flux,
under global boundary conditions of the type intoduced by Atiyah, Patodi and
Singer \cite{APS}. To this end, we have proceeded in two steps: In the first
place, we have grown the gauge field while keeping the boundary condition
fixed. This calculation was possible thanks to the exact knowledge of the zero
modes and the Green's function (in the complement
of the null space.) Here, a gauge invariant point splitting regularization was
employed. In references \cite{a1,a2} we developed, for the case of local
boundary
conditions, a regularization scheme based on Seeley's complex powers
\cite{Seeley}. Its application to the present problem leads to the same result
as the point splitting  we used here, even though its relation to the
$\zeta$-function is not guaranteed. In fact, as far as we know,  the
construction of complex powers for elliptic boundary problems with global
boundary conditions is still under study \cite{SG}.

In the second step, we have explicitly obtained the eigenvalues of  $ (i\not
\!\partial  +P_0)_{{\ka }}$. We have shown that the corresponding
$\zeta$-function is regular at the origin  and we have evaluated the quotient
of the free Dirac operators for two different global boundary conditions.

We have verified that our complete result is in agreement with the APS index
theorem, adapted to our example.

\bigskip
\bigskip
\bigskip
{\bf Acknowledgement}: We thank Mar\'\i a Amelia Muschietti and Jorge 
Solomin for valuable discussions.


\begin{thebibliography}{99}

\bibitem{r0}
M.~Bordag, B.~Geyer, K.~Kirsten and E.~Elizalde,
hep-th/9505157.


\bibitem{r1}
J.S. Dowker and J.S. Apps, hep-th/9506204.

\bibitem{r2}
E.~Elizalde, M.~Lygren and D.V. Vasilevich,
hep-th/9602113.

\bibitem{r3}
Klaus  Kirsten and Guido Cognola,
Class. Quant. Grav.{\bf 13}, 633 (1996).

\bibitem{r4}
Peter~D. D'Eath and Giampiero Esposito,
 Phys. Rev. {\bf D43}, 3234 (1991).

\bibitem{r5}
Peter~D. D'Eath and Giampiero Esposito,
 Phys. Rev. {\bf D44}, 1713 (1991).

\bibitem{r6}
A.~Wipf and S.~Durr,
 Nucl. Phys. {\bf B443}, 201 (1995).

\bibitem{r7}
M.~De Francia, H.~Falomir and E.~M. Santangelo,
Phys. Rev. {\bf D45}, 2129 (1992).

\bibitem{r8}
M.~De Francia,
Phys. Rev. {\bf D50}, 2908 (1994).


\bibitem{a1}
{H. Falomir,  R.E. Gamboa Sarav\'{\i}, M.A. Muschietti, E.M. Santangelo and
J.E.~Solomin, Determinants of  Dirac operators with local boundary conditions,
Journal of Mathematical Physics, in press.}

\bibitem{a2}
{H. Falomir,  R.E. Gamboa Sarav\'{\i}, M.A. Muschietti, E.M. Santangelo and
J.E.~Solomin, On the relation between determinants and Green functions of Dirac
operators with local boundary conditions, submitted to
Bulletin des Sciences Math\'ematiques.}

\bibitem{Seeley}R. T. Seeley,  Am. J. Math. {\bf 91}, 889 (1969); {\bf 91}, 963
(1969).

\bibitem{APS}
M.F. Atiyah, V.K. Patodi and  I.M. Singer, Math. Proc.
Camb. Phil. Soc. {\bf 77}, 43 (1975).

\bibitem{egh}
T. Eguchi, P. B. Gilkey and A. J. Hanson, Phys. Rep. {\bf 66}, 213 (1980)

\bibitem{Sitenko}
A. V. Mishchenko and Yu. A. Sitenko,
Ann. Phys. {\bf 218}, 199 (1992).


\bibitem{Moreno}E. F. Moreno,  Phys. Rev. {\bf D48}, 921 (1993).

\bibitem{Schwinger}J. Schwinger, Phys. Rev. {\bf 82}, 664 (1951);
{\bf 128}, 2425 (1962).




\bibitem{SG}G. Grubb and R.T. Seeley,  C.R. Acad. Sci. Paris, {\bf 317}, serie
I, 1123 (1993); Zeta and eta fuctions for Atiyah-Patodi-Singer
operators, Copenh. Univ., Math.  Dept, Preprint Ser., (1993), nro. 11; Weakly
parametric pseudodifferentials operators and
Atiyah-Patodi-Singer boundary ploblems, Copenh. Univ.,
Math. Dept, Preprint Ser., (1993), nro 15.



\bibitem{Tabla}  I.S. Gradshteyn and I.M. Ryzhik, {\em Table of Integrals,
Series, and Products}, Academic Press. Inc. (1980).


\bibitem{Annals}See, for example, R.E. Gamboa Sarav\'\i , M.A. Muschietti, F.A.
Schaposnik and J.E.Solomin,  Ann. Phys. {\bf 157}, 360 (1984).



\end{thebibliography}
\end{document}